1

# Modular Multilevel Converter with Sensorless Diode-Clamped Balancing through Level-Adjusted Phase-Shifted Modulation

N. Tashakor, *Student Member, IEEE*, M. Kilictas, E. Bagheri, and S. Goetz, *Member, IEEE*

*Abstract*—Cascaded H-bridge and modular multilevel converters (MMC) are on the rise with emerging applications in renewable energy generation, energy storage, and electric motor drives. However, their well-known advantages come at the price of complicated balancing, high-bandwidth isolated monitoring, and numerous sensors that can prevent MMCs from expanding into highly cost driven markets. Therefore, an obvious trend in research is developing control and topologies that depend less on measurements and benefit from simpler control. Diode-clamped topologies are considered among the more applicable solutions. The main problem with a diode-clamped topology is that it can only balance the module voltages of a string in one direction; therefore, it cannot provide a completely balanced operation. This paper proposes an effective balancing technique for the diode-clamped topology. The proposed solution exploits the dc component of the arm current by introducing a symmetrically level-adjusted phase-shifted modulation scheme, and ensures the balancing current flow is always in the correct direction. The main advantages of this method are sensorless operation, no added computation and control effort, and low overall cost. Analysis and detailed simulations provide insight into the operation of the system as well as the new balancing technique and the experimental results confirm the provided discussions.

## I. Introduction

MODULAR multilevel converters (MMCs) are the preferred voltage source inverter topology in most high-voltage applications, with high expectations for many new applications in the near future [1]. Modularity, scalability, high power quality, and flexibility are main features that distinguish MMCs from other multilevel converters [2]. With high expectations for MMCs, a large body of research focus on the barriers for MMCs further expansion, particularly in cost driven applications [3]. Imbalance of module voltages, complex and expensive monitoring systems, and complicated control algorithm are among the main hindrances for MMCs' expansion [4, 5].

Parameter spread, parasitics, discretization delay, or uneven use and associated aging of arm modules can lead to voltage imbalance in the modules and ultimately failure of the whole system [6]. The typical solution for the voltage balancing is sorting the modules from lowest to highest voltages and activate them in order [7-9]. However, tracking the module voltages is a prerequisite that can involve a costly monitoring system with an isolated high-bandwidth interface and computationally demanding algorithms [10-13].

Many have attempted to simplify the cell-sorting methods. Deng et al. propose to control the current of each module independently in a decentralized approach [14], while Wang et al. suggest a lower frequency sorting algorithm combined with a phase-shifted carrier (PSC) modulation [15]. Although such approaches are to some degrees successful in reducing the computation, they depend on voltage measurement and suffer from expensive and isolated monitoring requirement [16, 17].

Another research direction aims at reducing the cost of monitoring through online estimation of the module voltages. Kalman filter and sliding mode techniques are among the most popular choices [10, 18-20]. Generally, estimation methods trade a lower cost in the monitoring sub-system with even more computation. Additionally, susceptibility to model uncertainty and cumulative measurement errors can lead to divergence, even though accurate models may be successful in short periods [10]. Furthermore, in some applications balancing may be safety relevant to avoid failure or electric shock, which sets exceptional constraints on software (components of control).

Another class of voltage balancing techniques utilize cyclic switching patterns [21-24]. These techniques assume identical modules in each arm and then attempt to use a fixed-switching pattern to evenly distribute the load among them through cyclic permutations. However, they cannot guarantee convergence of the module voltages to a tight boundary when mismatches between module parameters exist and/or in case of an unpredictably fluctuation load.

Another alternative is modifying the topology of the MMC arm or modules to provide a balancing path with simpler control algorithms and lower sensory data [25-27]. Full-bridge or dual full-bridge with additional parallel mode between two modules are examples of such approaches that provide a balancing path via the parallel connection [11, 28-30]. Yet, the extra cost incurred by the higher number of individual semi-conductors and drivers as well as a more complex structure can prevent many of these topologies from utilization in high-voltage applications [31].

Diodes are inherently cheaper and simpler than their fully controlled counterparts. Therefore, diode-clamped topologies can be considered as the simpler and cheaper solution to a self-balancing MMC topology. Figure 1 shows the simplest form of a diode-clamped topology, even though approaches with higher complexity exist too [26, 32, 33]. In Fig. 1, the extra diode connects the positive terminals of two consecutive module capacitors and with the help of the main switches offers a unidirectional balancing path along a module string. However, since module excess energy can only move from a lower module to an upper one, imbalance persists when the upper module has a higher charge level.

To solve the unidirectional current flow limitation, Gao et al. use a high-voltage dc/dc converter that transfers energy from the first module in a string to the last [25]. However, a high-voltage switching transformer, which has to isolate the entire string voltage, defeats the purpose of utilizing the diode-clamped topologies to simplify balancing and reduce cost. Furthermore, the extra power conversion stages can reduce the balancing efficiency. In another solution, Gao et al. use two strings in parallel in each arm with opposite balancing directions, where each string shares half of the load current [26]. Although sharing the load reduces the cost of high-current semiconductors, the overall cost and complexity is still much higher than that of a normal MMC topology.

As a promising approach, Liu et al. propose a feedback control that measures the voltage of the top module in a diode-clamped topology ($u_{C_1}$) and control its modulation index to achieve a balanced state at all times [33]. A relatively similar approach is proposed in [34]. These methods use the simplest diode-clamped topology and a PID controller. However, these methods require constant voltage monitoring of the top module in each arm and adjust its target module voltage until it is lower than all others. However, since the top module is the only actuated one and by discharging them through the diode backbone instead of actively charging up negative outliers, it is limited by the weakest link and tends to suffer from large loss. As in conventional MMCs, voltage sensors here require high-voltage isolation, particularly since the first module is at the highest electrical potential in the arm.

This paper proposes a simple balancing technique for the diode-clamped topology through a so-called level-adjusted phase-shifted carrier (LAPSC) modulation. The proposed technique implements an open-loop control without any extra measurements.

The remainder of this paper is organized as follows: Section II studies the diode-clamped MMC topology. Section III explains the balancing principle of the proposed technique and discusses the balancing power loss. Section IV provides a general comparison between other state-of-the-art balancing techniques. Section V presents simulation and experiment results while Section VI concludes the paper.

## II. Principle Operation of a Diode-clamped Circuit

This section describes the working principle of the diode-clamped circuit depicted in Fig. 1. In addition to the conventional MMC circuit, $N - 1$ series clamping units connect the positive terminal of each module's dc-link to the next through a diode and an inductor.

### A. Clamping operation

Based on the numbering convention in Fig. 1, the $j^{th}$ clamping path connects the $j^{th}$ and $(j + 1)^{th}$ modules. The voltage across the diode depends on the control signal of the $(j + 1)^{th}$ switch. With $S_{(j+1)2}$: off and $S_{(j+1)1}$: on, diode $D_j$ is reverse biased with $-u_{C_j}$. However, when $S_{(j+1)2}$: on and $S_{(i+1)1}$: off, the voltage across diode $D_j$ is the voltage difference between $C_{j+1}$ and $C_j$ per

$$u_{D_j} = \begin{cases} -u_{C_j} & if\ S_{(j+1)2}: off \\ u_{C_{j+1}} - u_{C_j} & if\ S_{(j+1)2}: on \end{cases} \quad (1)$$

Hence, when $S_{(i+1)2}$ is on, if $u_{C_{j+1}} > u_{C_j} + V_{\text{fd}}$, the diode and the inductor can form a parallel connection between $C_j$ and $C_{j+1}$. The balancing current can flow from $C_{j+1}$ to $C_j$ through the clamping circuit as shown in Fig. 2(a). The role of $L_j$ is to protect the components from large current spikes by limiting the balancing current. When switch $S_{(j+1)2}$ is turned off again, the corresponding clamping diode is reverse biased and the inductor current decays with the rate of $\frac{di}{dt} = \frac{-(u_{C_j}+V_{\text{fd}})}{L_j}$, see Fig. 2(b). This procedure is repeated until $u_{C_{j+1}} \approx u_{C_j}$. As a result of suppressing the voltage difference, the inductor current stays zero. Figure 3 provides the voltage and current waveforms during the balancing operation.

Regardless of the initial values of $u_{C_j}$ and $u_{C_{j+1}}$, the final relation between the two capacitor voltages would be $u_{C_j} \geq u_{C_{j+1}}$. A similar analysis can be performed for the whole arm and

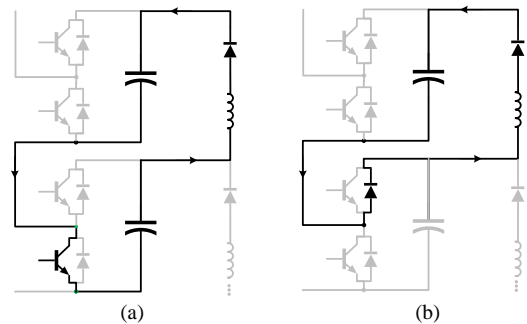

(a)        (b)
Fig. 2. Diode Clamping Current Path

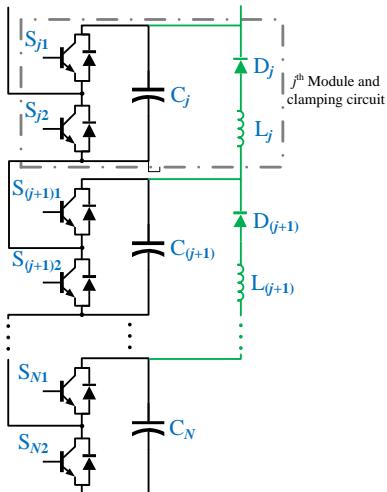

Fig. 1. Diode-clamped MMC Topology

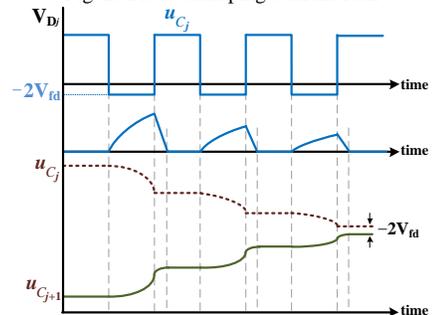

Fig. 3. A typical representation of the balancing process

the result follows
$$u_{C_1} \geq u_{C_2} \geq \cdots \geq u_{C_N}, \quad (2)$$
where $N$ is the total number of modules in one arm.

According to (2), balanced operation is ensured under the condition that the required balancing current flows from the bottom to the top of the arm. Therefore, the inequality relation in (2) becomes an equality neglecting the small voltage drop on the diodes ($u_{C_1} \approx u_{C_2} \approx \cdots \approx u_{C_N}$).

*B. Proposed level-adjusted phase-shifted carrier Modulation*

The PSC modulation compares a reference waveform with multiple phase-shifted carriers to generate control signals of the modules. In a conventional PSC modulation, each carrier corresponds to one module in the arm and the phase-shift between two consecutive carriers is $\frac{2\pi}{N}$. Neglecting non-ideal conditions, PSC can achieve a relatively stable operation in high-switching frequencies, which is investigated in [6]. However, as the switching frequency falls or as parasitics and mismatch between parameters increase, the system becomes more unstable and starts to gradually diverge from the intended operation point when the module voltages are not actively maintained within their operation boundaries [35].

A diode-clamped circuit can ensure that (2) is always maintained, but to change (2) to an equality, we must ensure that the imbalance always leans toward increasing the voltages of the lower modules. Therefore, the required balancing current would flow from bottom to the top of the arm. To that end, we introduce a small vertical displacement to the normally inline phase-shifted carriers, resulting in level-adjusted phase-shifted carrier modulation. Figure 4 shows an intuitive representation of the suggested level and phase shifts for upper and lower arms of each phase.

A negative vertical shift in a carrier waveform increases the duration that its corresponding module is connected in series, while a positive displacement increases the duration that the module is in the bypass state. Figure 5 provides a visual representation of how the displacement can affect the modules. Since, the arm current has a positive dc component, a negative displacement in a carrier leads to gradual increase in the voltage of the corresponding module. Similarly, a reverse effect is expected from a positive displacement in a carrier. Consequently, we can control the balancing direction and current by controlling the vertical-displacements of the carriers. To ensure the balancing direction is always from the bottom to the top of the arm, the displacements of carriers in one arm must follow
$$\delta_1 \geq \delta_2 \geq \cdots \geq \delta_j \geq \cdots \geq \delta_N, \quad (3)$$
where $\delta_j$ is the vertical-displacement of $j^{\text{th}}$ carrier.

The added displacement changes the modulation of each module slightly. The modulation of the $j^{\text{th}}$ module in the upper arm would be
$$m_j = \frac{1 - m_a \sin(\omega t)}{2} - \delta_j, \quad (4)$$
where $m_a$ is the modulation index of phase $a$.

Averaging the modulation waveforms of all the modules results in an effective arm modulation waveform per
$$m_p = \frac{\sum_j m_{pj}}{N} = \frac{1 - m_a \sin(\omega t)}{2} + \frac{\sum_j^N \delta_j}{N}. \quad (5)$$

Therefore, for identical output as a normal MMC, the second term in (5) should be equal to zero ($\sum_j^N \delta_j = 0$), resulting in

$$m_p = \frac{1 - m_a \sin(\omega t)}{2}. \quad (6)$$

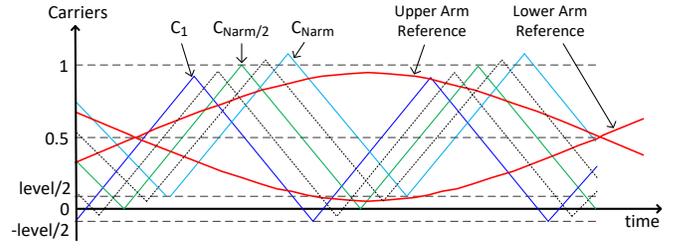

Fig. 4. Module carriers with displacement

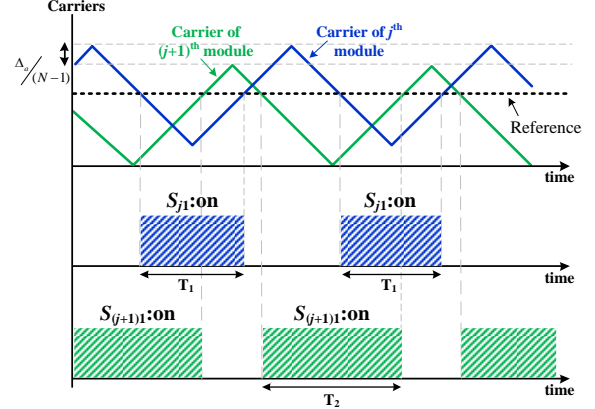

Fig. 5. Effect of displacement on the module states

According to (3) and (6), we can ensure a balanced operation in the arm by controlling $\delta_j$, while keeping the output identical to a normal MMC. Hence, the added displacements should be set following
$$\begin{cases} \delta_1 \geq \delta_2 \geq \cdots \geq \delta_j \geq \cdots \geq \delta_N, \\ \sum_j^N \delta_j = 0, \end{cases} \quad (7)$$

We propose $\delta_j$ to be calculated with respect to the total displacement value per
$$\delta_j = \Delta_a \left( \frac{1}{2} - \frac{j-1}{N-1} \right). \quad (8)$$
where $\Delta_a$ is the total displacement between the first and last carriers of an arm in phase $a$.

Defining displacements according to (8) increase the overall symmetry since the displacement difference between every two neighboring modules is $\frac{\Delta_a}{N-1}$. Furthermore, (8) satisfies the conditions of (7).

The procedure for the lower arm is identical; however, the phase-shift orders of the carriers in the upper and the lower arm are reversed, i.e., if the $\boldsymbol{\phi}_{upper} = \left[0, \frac{\pi}{N}, \frac{2\pi}{N}, \ldots, \frac{2\pi(N-1)}{N}\right]^T$ is the vector of the carrier phase shifts in the upper arm, the vector of the carrier phase shifts in the lower arm is $\boldsymbol{\phi}_{lower} = \left[\frac{2\pi(N-1)}{N}, \frac{2\pi(N-2)}{N}, \ldots, 0\right]^T$. It would ensure that the total displacement of the modules connected at each instance is also zero.

Because of the existing symmetry in the carrier displacements of one arm, we can only analyze the required displacement to balance the first and last modules in the arm (similarly, the second and the one before the last module can be balanced together and so on). Considering a mismatch between the capacitances of the first and last modules, the increase of voltage difference in one switching cycle when $C_N > C_1$ is calculated per
$$\Delta V = \frac{1}{2f_1} \left( \frac{1}{C_1} - \frac{1}{C_N} \right) I_{imbalance}, \quad (10)$$

where $f_1$ is the fundamental frequency of the output voltage $I_{imbalance}$ is the imbalance current caused by parameter mismatch and/or discretization delay [6].

It is possible to write the capacitance values according to the rated capacitance ($C$) of the modules as $C_1 = \left(1 - \frac{N-1}{2}\varepsilon\right)C$ and $C_N = \left(1 + \frac{N-1}{2}\varepsilon\right)C$, where $\varepsilon$ depends on the average tolerance of the module capacitances throughout the arm, e.g. with the maximum tolerance of $\pm 15\%$, $\varepsilon = \frac{0.3}{N-1}$. Substituting the capacitance values in (10) results in

$$\Delta V_{1,N} = \frac{1}{2f_1}\left(\frac{1}{1-\frac{N-1}{2}\varepsilon} - \frac{1}{1+\frac{N-1}{2}\varepsilon}\right)I_{imbalace}. \quad (11)$$

The exact value of maximum imbalance current can be only derived through numeric methods. However since it depends on the module mismatch, system parameters, and arm current, we can approximate it by $I_{imbalance} \approx \varepsilon I_{dc}$. Assuming lossless analysis, the upper arm current can be written based on the phase current as

$$i_{arm,p} = \frac{I_P}{2}\left(\frac{1}{k} + \sin(\omega t - \varphi)\right), \quad (12)$$

where $I_P$ is the amplitude of the phase current, $\varphi$ is the load angle, and $k = \frac{2}{m_a \cos\varphi}$.

Substituting the dc component of the arm current in (11) and after some mathematical manipulation, the average voltage difference after one cycle of output voltage emerges as

$$\Delta V_{1,N} = \frac{I_P}{4kf_1C}\frac{(N-1)\varepsilon}{(1-\frac{N-1}{2}\varepsilon)(1+\frac{N-1}{2}\varepsilon)}. \quad (13)$$

The compensated voltage difference between first and last modules due to the added displacement is

$$\Delta V_{\Delta_a} = \frac{I_P \Delta_a}{4kf_1C}\left(\frac{2}{(1-\frac{N-1}{2}\varepsilon)(1+\frac{N-1}{2}\varepsilon)}\right). \quad (14)$$

As long as $\Delta V_{\Delta_a} \geq \Delta V_{1,N}$, the clamping circuit will be able to keep the system in balance. Therefore, the minimum required displacement for balancing capacitor mismatches is

$$\Delta_a > \frac{(N-1)\varepsilon^2}{2}. \quad (15)$$

The actual displacement, should be slightly higher to also account for the discretization delay [6].

### III. DESIGN CONSIDERATIONS AND SYSTEM ANALYSIS

This section analyzes the clamping circuit, provides guidelines for the component selection, and investigates the balancing loss.

#### A. Circuit Analysis

With a parallel connection between the two modules as shown in Fig. 2, the equivalent electrical circuit is a series RLC circuit, where $C_e = \frac{1}{2}C_j = \frac{1}{2}C_{j+1}$, $u_{diff} = u_{C_{j+1}} - u_{C_j} - V_{fd} - V_{sw}$, and $R = 2r_C + r_d + r_S + r_L$. The second-order differential equation governing this circuit is

$$\frac{d^2 u_{diff}}{dt^2} + \frac{R}{L_j}\frac{du_{diff}}{dt} + \frac{1}{L_j C_e}u_{diff} = 0. \quad (16)$$

Solving the second-order equation leads to two roots,

$$P_{1,2} = \frac{-R}{2L_j} \pm \sqrt{\left(\frac{R}{2L_j}\right)^2 - \frac{1}{L_j C_e}}. \quad (17)$$

The equivalent resistance of the balancing path is relatively small ($R < 2\sqrt{\frac{L_j}{C_e}}$) and the system should demonstrate damped oscillations with damped frequency $\sqrt{\frac{1}{L_j C_e} - \left(\frac{R}{2L_j}\right)^2}$. However, because of the existance of diode, no oscillations occur in the clamping current because the inductor current can not be negative. When switch $S_{(j+1)2}$ is on, the voltage difference charges the inductor with a rate of $\frac{u_{diff}}{L_j}$, and when $S_{(j+1)2}$ is turned-off, the inductor discharges into $C_j$ with a rate of $\frac{-u_{C_j} - 2V_{fd}}{L_j}$, until it reaches zero. In normal operation, $u_{C_j} \gg u_{diff}$ and the inductor current decays to zero almost imidiately. When $S_{(j+1)2}$ is turned-on, the balancing current is

$$i(t) = \begin{cases} Ae^{-\alpha t}\sin(\omega_d t), & 0 \leq \omega_d t \leq \pi \\ 0, & \omega_d t > \pi \end{cases}. \quad (18)$$

where $\alpha = \frac{R}{2L_j}$, $\omega_0 = \frac{1}{\sqrt{L_j C_e}}$, $\omega_d = \sqrt{\omega_0^2 - \alpha^2}$, $i(0) = 0$, $\frac{di(0)}{dt} = \frac{u_{diff}}{L_j}$, and $A = \frac{u_{diff}}{\sqrt{\frac{L_j}{C_e} - \frac{R^2}{4} - \frac{R}{2}}}$. Therefore, the peak value that the inductor current can reach is estimated using

$$I_{peak,1} < \frac{U_{diff,max}}{\sqrt{\frac{L_j}{C_e} - \frac{R^2}{4} - \frac{R}{2}}}, \quad (19)$$

where $U_{diff,max}$ is the maximum permissible voltage difference. However, when $\frac{1}{\omega_d}\tan^{-1}\frac{\omega_d}{\alpha} < DT_{sw}$, the inductor current never reaches its peak value (because $S_{(j+1)2}$ is turned-off) and the maximum inductor current is approximated according to

$$I_{peak,2} < Ae^{-\alpha D_{max}T_{sw}}\sin(\omega_d D_{max}T_{sw}), \quad (20)$$

where $0 \leq D_{max} \leq 1$ is the maximum duty cycle of $S_{(j+1)2}$, and $T_{sw}$ is the switching cycle. In the worst-case scenario, $D_{max}$ is equal to one.

The peak value of the inductor current determines the current ratings of diode $D_j$ [33]. Therefore, selecting the maximum current rating of $D_j$, the clamping inductor value follows

$$L_j \geq \min\left(\left[\left(\frac{U_{diff,max}}{I_{D,max}} + \frac{R}{2}\right)^2 + \frac{R^2}{4}\right]C_e, \frac{U_{diff,max}T_{sw}}{I_{D,max}}\right). \quad (21)$$

Smaller inductor values increase the speed of balancing at the cost of a higher diode current, whereas a larger inductor reduces the speed of balancing, but allows for a smaller diode [31].

Equation (19) limits the lower boundary of the inductance, while the displacement as well as switching and fundamental frequencies of the system bound the upper limit. The inductor should be low enough that it can balance the added displacement current in one cycle of the arm current. The amplitude of the average balancing current based on the displacement is

$$\bar{I}_{\Delta_a} = I_{dc}\Delta_a = \frac{I_P \Delta_a}{2k}. \quad (22)$$

The modules are balanced when $S_{(j+1)2}$ is on and the average duration that $S_{(j+1)2}$ is on follows

$$\bar{T}_{S_{(j+1)2}} = \frac{0.5}{f_{sw}}. \quad (23)$$

The average achievable current of the inductor should be higher than the average displacement current per

$$\frac{U_{diff,max}}{L_i}\frac{\bar{T}_{S_{(j+1)2}}}{2} > \frac{I_P \Delta_a}{2k}, \quad (24)$$

and the upper baundary of the clamping inductance is

$$L_i \leq \frac{\Delta_a \times I_P}{\bar{T}_{S_{(j+1)2}} U_{diff,max}}. \quad (25)$$

Therefore, after determination of the maximum permissible diode current, equations (19) and (23) determine the required inductor value.

*B. Power Loss Analysis*

This section investigates the power losses and provides some insight into the balancing loss due to the added displacement. In Section II, we showed that the displacement does not change the system behavior in the arm and phase level. Furthermore, the total number of switches/diodes that are conducting throughout the arm remain constant because the displacements and phase-shifts are defined in a complementary manner, see Fig. 3. Therefore, we can estimate the power loss due to the arm current similar to conventional MMCs using numerical methods and independent of the displacement value [36]. Alternatively, we can derive simpler approximations based on the average and RMS values of the arm current, assuming identical conduction loss for power switches and diodes (i.e., $V_{sw} = V_{fd} = V_0$ and $r_{sw} = r_d = r$).

The switch–diode RMS current, capacitor RMS current, and the switch/diode average current are

$$I_{RMS,cap} = \frac{I_P}{4}\sqrt{\frac{m_a^2+2-4km_a\cos(\varphi)}{2k^2} + \frac{4m_a^2+4-m^2\cos(\varphi)}{8}}, \quad (26)$$

$$I_{RMS,arm} = \frac{I_P}{2}\sqrt{\frac{1}{k^2}+\frac{1}{2}}, \quad (27)$$

$$I_{avg,arm} = I_{dc} = \frac{I_P}{2k}. \quad (28)$$

The power loss in one arm is

$$P_{loss} = NI_{avg}V_0 + NrI_{RMS,arm}^2 + Nr_cI_{RMS,cap}^2 + \underbrace{2Nf_{sw}}_{N_{sw}}\left(\frac{1}{2}V_mI_{avg,sw}(t_{on}+t_{off})\right), \quad (29)$$

where $t_{on}$ and $t_{off}$ are turn-on and turn-off durations, and $V_m$ is the nominal voltage of one module.

The power loss calculated in (27) does not include the balancing loss, and one should calculate it separately. In general, the exact value of the balancing loss can only be calculated using detailed numerical simulations. However, it is possible to approximate the maximum additional power loss.

The added displacement generates a circulating current between the modules. Using the symmetry between the module displacements, the extra power loss due to the added circulating current between $j^{th}$ and $(N-j)^{th}$ modules is

$$E_{loss}^{j,N-j} = \left(\frac{I_{RMS,arm}\Delta_a}{(N-1)}\right)^2(N-2j+1)^2(r_L+2r) + \left(\frac{I_{RMS,arm}\Delta_a}{(N-1)}\right)(N-2j+1)V_0. \quad (30)$$

The total balancing loss of one arm is summation of the energy lost between all the modules and is calculated as

$$P_{balancing} = \left(\frac{I_{RMS,arm}\Delta_a}{(N-1)}\right)^2(r_L+2r)\sum_{j=1}^{\frac{N}{2}}(N-2j+1)^2 + \left(\frac{I_{RMS,arm}\Delta_a}{(N-1)}\right)V_0\sum_{j=1}^{\frac{N}{2}}(N-2j+1). \quad (31)$$

Based on (29), the balancing loss increase as load current and/or displacement increase.

## IV. GENERAL COMPARISON

Table I presents an overview of the balancing solutions based on topology modification and compares them with the proposed method. The main advantages of the proposed solution include the following:
- with only one extra diode and sensorless balancing, the proposed topology has the minimum extra components;
- the topology can achieve stable operation without any control requirement for balancing the module voltages;
- the proposed method can achieve a good efficiency, which the presented power loss analysis in Section III-B and the simulation results in Section V confirm;
- the proposed balancing has no adverse effect on the output of the system.

## V. SIMULATION AND EXPERIMENTAL RESULTS

We analyze the proposed LPSC modulation technique with a single-phase model in MATLAB/Simulink. Additionally, a prototype with eight modules provides a proof of concept. Table II lists the parameters of the simulation and the experimental systems. The semiconductors in the simulation are modelled after SEMiX854GB176HDs power modules from Semikron. Other system parameters are determined based on the discussion in Sections II and III.

TABLE I
GENERAL OVERVIEW OF THE EXISTING BALANCING METHODS

| Solution | Ref | Switches | Diodes | Inductors | Comments | Advantages | Disadvantages |
|---|---|---|---|---|---|---|---|
| Diode-clamped | [25] | 2n+4 | n+3 | 0 | Achieves bi-directional balancing by a high-voltage transformer connecting bottom and top modules together | Low sensitivity to parameter variations; lower voltage sensors | Multiple power conversion stages affect the efficiency; higher cost due to the extra transformer |
| Dual string diode-clamped | [26] | 4n | 2n | 2n | Two separate strings are connected in parallel to form an arm, where each string has a different balancing direction | Sharing the power between strings; sensorless operation | High number of components increases the cost |
| MMSPC | [27] | 4n-4 | 0 | 2n-2 | Provides parallel connection for the modules and realizes balancing through parallel functionality | Four-quadrant operation; improved efficiency; sensorless operation | Two extra switches and one additional capacitor increase the cost |
| Two parallel clamping clusters | [31] | 3n-1 | 2n-2 | 2n-2 | A diode-clamped and a switch-clamped balancing path in parallel | Sensorless operation; simple control | Lower Efficiency; medium number of components |
| Diode-clamped | [37, 38] | 2n | n-1 | n-1 | Closed-loop control of upper module's modulation reference | Fast SM voltage balancing | Requires measurement of top module voltage; extra diodes |
| Switch-clamped | [35] | 3n-1 | 0 | n-1 | Switch-clamped balancing path between modules as well as phases | Sensorless operation | Medium number of additional components |
| Double-clamped | [39] | 3n | 2n-2 | 2n-2 | Double diode-clamped circuit for balancing | Simple balancing strategy; sensorless operation | High number of components, double diode-clamped structure can lead to ringing effect |
| Double-Star Submodule | [40] | 10n | 2n | 0 | Fault tolerant operation by combining two modules into one | DC fault blocking capability | Requires voltage sensor for each submodule; high number of extra components; complex structure and complicated control |
| Proposed Method | | 2n | n-1 | n-1 | Open-loop diode-clamped balancing using the dc component of the arm | Sensorless operation; simple control; few extra components; high efficiency | The convergence speed can be slow |

TABLE II
SINGLE-PHASE SIMULATION AND EXPERIMENT PARAMETERS

| Circuit Parameters | Simulation | Experiment |
|---|---|---|
| Number of SMs | 40 | 8 |
| dc voltage | 24 kV | 120 V |
| dc Capacitor | 15 mF | 4.9 mF |
| Grid frequency | 50 Hz | 50 Hz |
| Modulation index ($m$) | 0.95 | 0.95 |
| Switching frequency | 5 kHz | 10 kHz |
| Arm inductor | 10 mH | 2 mH |
| Clamping inductor | 10 µH | 7.5 µH |

*A. Simulation*

We consider four different systems: $i$) identical modules with $\Delta_a = 0$, $ii$) identical modules with $\Delta_a = 0.002$; $iii$) mismatch between module capacitors; $iv$) modules with different self-discharge rates. The capacitances and internal resistances of the mismatched modules in the third system are defined per $C_j = \left(1.3 - 0.6 \times \frac{N-j}{N-1}\right) C$ and $r_{C_j} = \left(0.7 + 0.6 \times \frac{N-j}{N-1}\right)$. Also, Table III shows the modified parameters of the modules with higher self-discharge rate.

Figure 6 presents the output voltage of the MMC for different scenarios. Generally, all the phase voltages are completely sinusoidal with negligible differences. However, after zooming in, one can detect small high-frequency ripples due to the balancing operation. Additionally, Table IV shows the total harmonic distortion values for different scenarios. It demonstrates that even displacements as high as 2 % ($\Delta_a = 0.02$) have negligible negative effect (less than 0.04 %) on the output voltage.

Figure 7 investigates the balancing performance of the proposed technique. As analyzed in Section II, the simulation results in Fig. 7(a) show that even with identical modules, the voltages gradually diverge from their rated value. Figure 7(b) shows that the same system will be stable with $\Delta_a = 0.002$. According to the loss analysis, the maximum increase in power loss with $\Delta_a = 0.002$ is around 0.01 % in all conditions and according to Fig. 9, the increased power loss is even less.

Figure 8(a) shows the behavior of a system with modules that have different capacitances and internal resistances. At $t = 5\ s$, the displacement is increased from 0 % to 2 % and the module voltages start to converge immediately toward the rated value. We repeated the simulation with the same system while reducing the displacement each time; the system is able to return the voltage of the modules to nominal value with displacement values higher than 0.9 %, which confirms the analysis in Section II.

TABLE III
SINGLE-PHASE MODIFIED SIMULATION AND EXPERIMENT PARAMETERS

| Modified Simulation Parameters | Modification |
|---|---|
| Parallel resistance with $SM_{4,U}$ | 32 kΩ |
| Parallel resistance with $SM_{9,U}$ | 28 kΩ |
| Parallel resistance with $SM_{14,U}$ | 24 kΩ |
| Parallel resistance with $SM_{19,U}$ | 20 kΩ |
| Parallel resistance with $SM_{4,L}$ | 16 kΩ |
| Parallel resistance with $SM_{9,L}$ | 12 kΩ |
| Parallel resistance with $SM_{14,L}$ | 8 kΩ |
| Parallel resistance with $SM_{19,L}$ | 4 kΩ |
| Modified Experiment Parameters | Modification |
| Capacitors of $SM_{1,U}$ and $SM_{3,L}$ | 2.2 mF |
| Parallel resistance with $SM_{1,U}$ | 68 kΩ |
| Parallel resistance with $SM_{3,L}$ | 4.5 kΩ |

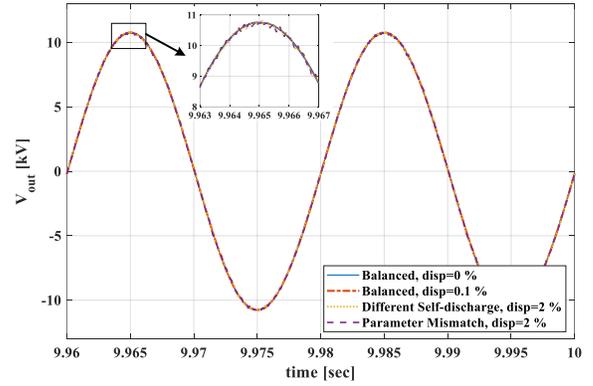

Fig. 6. Output voltages in different scenarios

TABLE IV
THD VALUES OF THE VOLTAGE

| Condition | THD$_V$ |
|---|---|
| Identical modules, $\Delta_a = 0$ % | 0.74 % |
| Identical modules, $\Delta_a = 0.1$ % | 0.74 % |
| Identical modules, $\Delta_a = 2$ % | 0.77 % |
| Modules with different capacitances, $\Delta_a = 2$ % | 0.77 % |
| Modules with different Self-discharge, $\Delta_a = 2$ % | 0.78 % |

However, the convergence speed is further decreased as we reduce the displacement. In a real application, the convergence speed does not present an issue since the system is always operated with the constant displacement. In the last scenario, we changed self-discharge rate of some of the modules to simulate different leakage and aging for different modules. Fig. 8(b) shows the module voltages during this scenario. Although there is severe mismatch between the modules, all the voltages start to converge to the rated value at $t = 7$ s, when we increase the displacement from 0 % to 2 %.

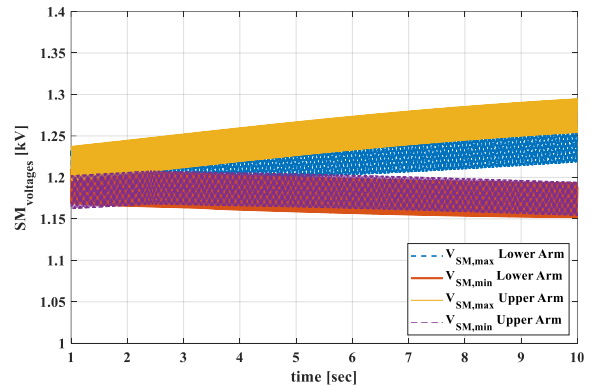

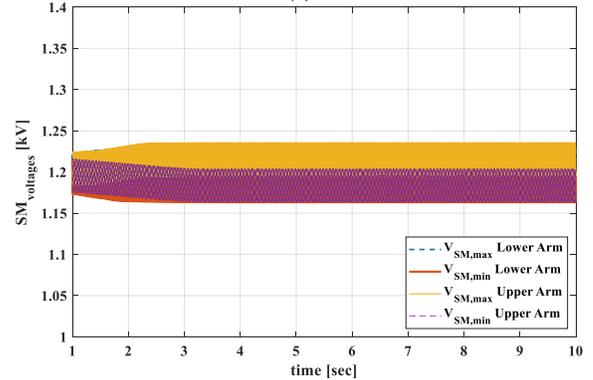

Fig. 7. Module voltages with balanced parameters: (a) $\Delta_a = 0$; (b) $\Delta_a = 0.002$

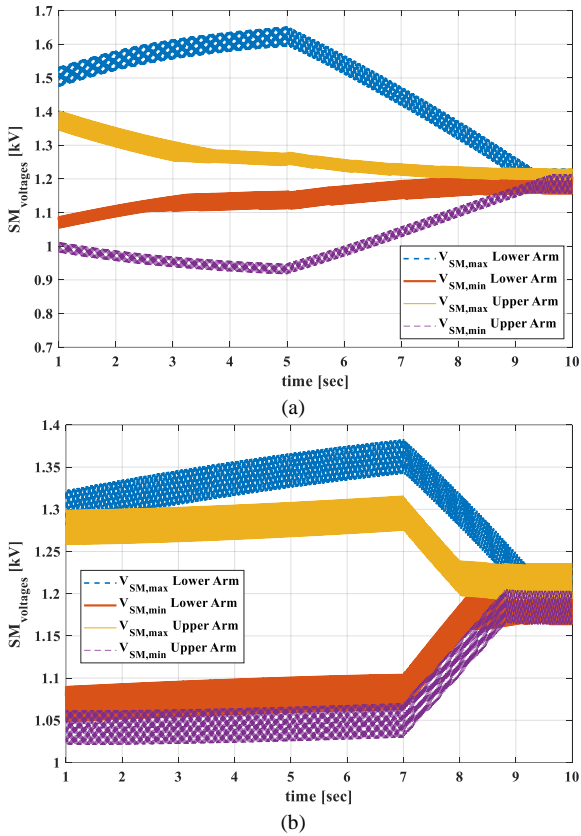

Fig. 8. Module voltages with imbalanced modules: (a) mismatch between the modules' capacitors, $\Delta_a = 0.02$; (b) different capacitance and self-discharge rates, $\Delta_a = 0.02$.

Figure 9 shows the power loss of a balanced system while the displacement keeps rising. According to the analysis in the last two sections, the minimum required displacement value in a real-life application is approximately 0.001 to 0.003 and based on Fig. 9, the added power loss is around 0.01 %.

*B. Experiment*

The parameters of the testbench are listed in Table II. Equations (19) and (23) limit the clamping inductor range to 3 μH − 10 μH. Passing a wire through a toroidal ferri-magnetic core results in approximately 7 μH with negligible resistance, which is within the calculated baundary [41]. The modules include low ESR ceramics and around 4.5 mF electrolytes. Labview in combination with an FPGA development board from National Instruments control the system and generate the switching pulses.

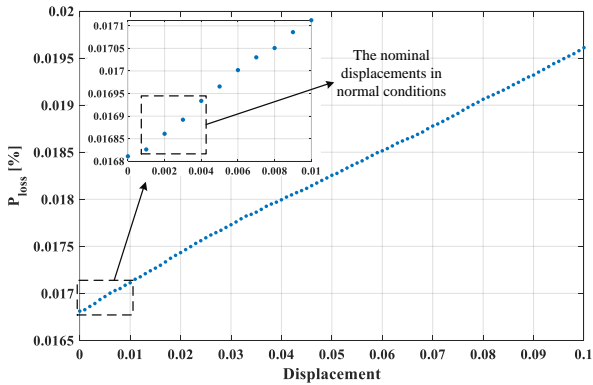

Fig. 9. Power loss in different scenarios

PSC modulation can achieve a five-level output voltage with four modules in each arm. Figure 10 shows the output voltages and currents for a balanced system as well as a severely imbalanced one with $\Delta_a = 0.02$. There are no discernible differences between the output of a system with identical modules (in Fig. 10(a)) and the output of a system with severly mismatched modules (in Fig. 10(b)). Table III lists the modified parameters of the modules for the second condition.

Furthermore, Fig. 11(a) shows the balancing performance with modules that are identical but have initial voltage imbalance. The voltages converge to the rated value after starting the system with a $\Delta_a = 0.02$. The balancing operation with a voltage spread of 50 % lasts less than 250 ms. Increasing the displacement resulted in faster convergence and the minimum dispositon that resulted in a continuously stable operation was 0.3 %. Figure 11(b) shows the voltages of the arm modules for a system with mismatched modules. Although the convergence is understandably slower, the system can achieve balanced operation with $\Delta_a = 0.02$ and the maximum voltage difference is limited to less than 1.5 %. A higher displacement will result in faster convergence, but it does not affect the maximum voltage difference. Only the switching frequency and the degree of imbalance among the modules determine the maximum voltage difference.

## VI. CONCLUSION

This paper proposes a simple and efficient balancing solution for a diode-clamped MMC topology. It introduces a level- and phase-shifted carrier modulation, which uses the dc component of the arm current to achieve balancing. Through analysis, simulation, and experiments, we confirm that the proposed method has a negligible effect on the ouput and can achieve open-loop sensorless operation. Furthermore, we demonstrate that the proposed method benefits from relatively low balancing loss. Based on simulations and experiments, the proposed method can maintain the module voltages of an imbalanced system within a 3 % boundary in case of simulation and 1.5 % boundary in case of experiments.The results suggest that the convergence speed is dependent on the degree of imbalance as well as the displacement value. However, a relatively low displacement value (in most cases less than 0.3 %) can prevent any imbalance accumulation.

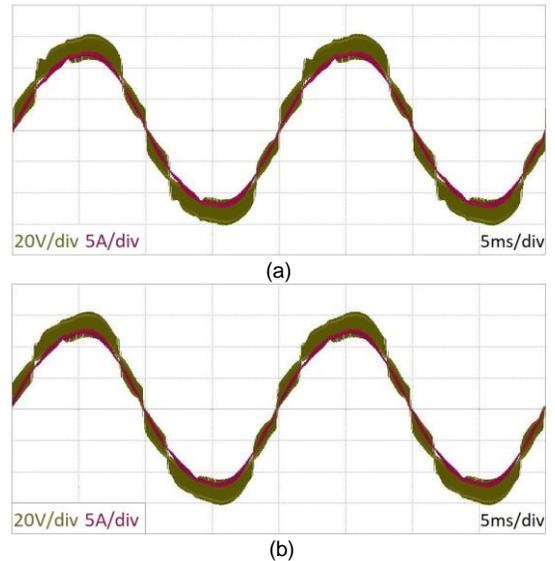

Fig. 10. Output voltage and output current of the lab prototype with the clamping circuit: (a) normal operation (b) unbalanced situation

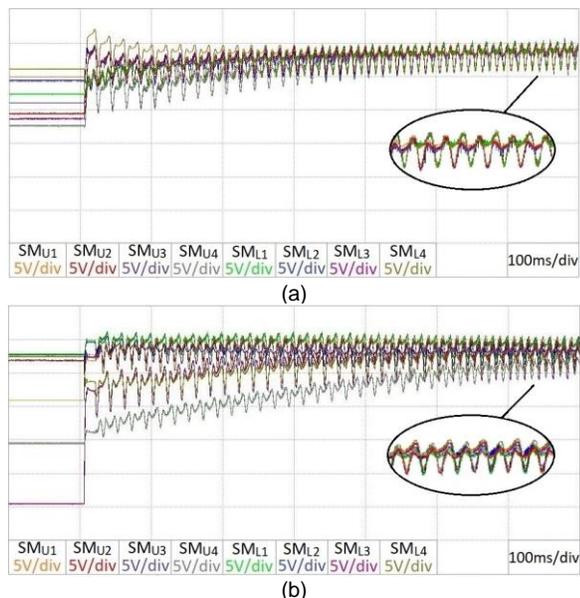

Fig. 11. Experimental Results: (a) module voltages with identical modules but initial voltage imbalance; (b) module voltages with severely mismatched modules